\documentclass[a4paper,fleqn,usenatbib]{mnras}

\usepackage[T1]{fontenc}
\usepackage{ae,aecompl}

\usepackage{graphicx}   
\usepackage{amsmath}    
\usepackage{amssymb}    
\usepackage{dblfloatfix}

\newcommand{\etal}{et al.}

\def\lesssim{\mathrel{\hbox{\rlap{\hbox{\lower4pt\hbox{$\sim$}}}\hbox{$<$}}}}
\def\gtrsim{\mathrel{\hbox{\rlap{\hbox{\lower4pt\hbox{$\sim$}}}\hbox{$>$}}}}
\def\apj{ApJ}
\def\apjs{ApJS}
\def\aj{AJ}
\def\aap{A\&\hskip-1pt A}
\def\aaps{A\&\hskip-1pt AS}
\def\mnras{MNRAS}
\def\pasp{PASP}

\def\jkas{JKAS}

\title[SFHs of dS0s]{Star Formation Histories of Dwarf Lenticular Galaxies}
\author[Ann \& Seo]{
Hong Bae Ann$^1$\thanks{E-mail: hbann@pusan.ac.kr} and Mira Seo$^1$ 
\\
$^{1}$Department of Earth Science Education, Pusan National University, 46241, Busan, Republic of Korea\\
}
\date{Accepted XXX. Received YYY; in original form ZZZ}

\pubyear{2022}

\begin{document}
\label{firstpage}
\pagerange{\pageref{firstpage}--\pageref{lastpage}}
\maketitle



\begin{abstract}

Star formation histories (SFHs) are essential for understanding galaxy formation and evolution.  We present the mean SFHs of 148 dwarf lenticular galaxies (dS0s) derived from the Sloan Digital Sky Survey (SDSS) spectra. These SFHs reveal two distinct periods of active star formation. The first period, concluding approximately 6 Gyr ago, witnessed the formation of predominantly old, metal-poor (Z=0.0004) stars, accounting for approximately 60\% of the stellar mass and 30\% of the luminosity. The scarcity of extremely metal-poor (Z=0.0001) stars suggests pre-enrichment during the re-ionization era. Star formation gradually waned during this initial phase. In contrast, the second period, ending around 1 Gyr ago, exhibited a peak in the middle of the period, contributing to the formation of moderately old stellar populations with intermediate metallicity. The SFHs of dS0 galaxies show a clear dependence on stellar mass, with more massive dS0s forming stars earlier. But, we find no significant correlation with morphological properties such as outer spiral arms and nucleation. The SFHs of dS0 galaxies share many similarities with those of dE galaxies, suggesting a common origin, mostly not primordial.

\end{abstract}

\begin{keywords}
 galaxies: dwarfs -- galaxies: formation -- galaxies: evolution -- galaxies: star formation 
\end{keywords}

\section{Introduction}
Dwarf lenticular galaxies (dS0s) are early-type dwarf galaxies whose morphology is similar to dwarf elliptical galaxies (dEs), but they possess a bulge/lens-like component in the inner regions. They were introduced by \citet{san84} as a class of dwarf galaxies in the Virgo Cluster. \citet{bin91} conducted a detailed examination of the morphologies of dS0s, using 20 dS0s in the Virgo cluster. Their study revealed a variety of disc features present in the two-dimensional images of dS0s. The surface brightness profiles of dS0s resemble those of dEs, which are well fitted by exponential profiles or King profiles, with an excess of light in the central regions \citep{bin91, bin93}. The similarity in surface brightness profiles of dEs and dS0s has raised questions regarding whether dS0s truly represent a distinct class of dwarf galaxies \citep[e.g.,][]{rey99}, leading to some researchers treating dS0s as a subclass of dEs \citep[e.g.,][]{bara03}.

However, if we accept the definition of dE and dS0 galaxies as introduced by \citet{san84}, it seems reasonable to differentiate dS0s from dEs based on their surface brightness profiles, as dS0s exhibit multiple components with one for the inner regions and another for the outer regions. The surface brightness profile of the inner component is most accurately described by the S\'{e}rsic profile \citep{ser68}, while that of the outer component conforms to an exponential profile. The inner component is generally regarded as a lens \citep{buta13}, whereas the outer component is seen as a disc. Leveraging this characteristic of dS0s' surface brightness, \citet{agu05} classified the early-type dwarfs in the Coma cluster into dEs and dS0s while \citet{jan12, jan14} classified the early-type dwarfs in the Virgo cluster using GALFIT \citep{pen10}, a two-dimensional decomposition tool. Their classification revealed significantly smaller dS0/dE ratios in the Coma cluster compared to the Virgo cluster. The reason for the higher fraction of dS0s in the Virgo cluster seems to be the better resolution owing to the shorter distance to the Virgo cluster. Additionally, the use of two-dimensional images rather than one-dimensional surface brightness profiles employed by \citet{agu05} may have contributed to this classification difference. 

Despite the examples of dS0s provided by \citet{bin91} and the effectiveness of two-component models demonstrated by \citet{agu05}, the ongoing debate revolves around whether dS0 should be considered a distinct, coequal category among early-type dwarfs. Many researchers prefer to classify all dwarf elliptical-like galaxies, including dwarf spheroidals (dSphs) and dwarf lenticulars (dS0s), under the broader category of dE galaxies \citep[e.g.,][]{bara03}. This tendency arises partly due to the limited number of dS0s in comparison to dEs, especially within the Virgo cluster, and the presence of disc features such as spiral arms, bars, and lenses observed in both dEs and dS0s \citep{jer00, jer01, bara02, rij03, grag03, gra03, fer06, lis06, lis09, jan12, pen14, gra17}.

The importance of morphological classification for dS0 galaxies has been reaffirmed in recent visual galaxy classification efforts, exemplified by projects such as the Extended Virgo Cluster Catalog \citep[][hereafter EVCC]{kim14} and the catalog of visually classified galaxies in the local universe within $z = 0.01$ \citep[][hereafter CVCG]{ann15}. Notably, the number of dS0s in the EVCC is nearly double that of the Virgo Cluster Catalog (VCC), while the count of dEs remains relatively stable. This increase in the number of dS0s in the EVCC primarily results from reclassifying certain dEs from the VCC as dS0s in the EVCC. Additionally, a slight decrease in the count of dEs in the EVCC is due to the reclassification of some dEs from the VCC as E, S0, and Sa galaxies.

The origin of dS0s is a subject of particular interest. Similar to dEs, there are two plausible scenarios for their formation: one related to their primordial origins and the other involving transformation processes. If dS0s are indeed transformed from late-type galaxies, the conversion of gas-rich disc galaxies into gas-poor early-type dwarfs likely involves two distinct phases. The initial phase entails gas removal, while the subsequent stage includes processes such as tidal heating and harassment. Given that a significant portion of dS0s are found in group and cluster environments \citep{mak11}, these transformation mechanisms may involve ram pressure stripping \citep{gg72} and galaxy harassment \citep{moo96, moo98}.

In contrast, \citet{bos08} and \citet{ste20} have reported that the transformation of late-type galaxies into early-type dwarfs can be achieved solely through ram pressure stripping, while \citet{chi09} suggested the possibility of major/minor mergers playing a role. The morphological transformation from late-type galaxies to early-type dwarfs is feasible because ram pressure not only removes cold gas but also compresses it, leading to star formation. The grand-design spiral arms observed in unsharp masked images of some early-type galaxies \citep[e.g.,][]{lis06} are believed to form in the gas compressed by ram pressure. The disc features, including arms, bars, lenses, and clumps, seen in the residual images of early-type dwarfs, especially dS0s, could result from star formation driven by ram pressure. Ram pressure has the effect of heating a thin stellar disc, turning it into a thicker one by removing the gas potential \citep{smi12}. Additionally, gas clouds pushed out of the disc by ram pressure can remain gravitationally bound, eventually falling back and fueling star formation in the thick disc \citep{ste20, bos22}.

Conversely, the hypothesis of a primordial origin for dS0s remains a viable one, supported by the fact that approximately 7\% of them exist as isolated galaxies \citep{ann24}, unaffected by environmental factors that could potentially alter their morphology. However, recent analyses of the star formation histories (SFHs) of dEs and dSphs \citep{seo23}, derived from the SDSS spectra, have revealed that a significant fraction of dEs cannot be considered primordial objects, as they exhibit active periods of star formation around 2.5 Gyr ago, whereas the majority of dSphs are believed to be primordial objects. Since dS0s share many similarities with dEs, it is reasonable to expect that the SFHs of dS0s are similar to those of dEs.

There have been relatively few studies on the SFHs of dS0s, primarily due to the scarcity of dS0s within the Local Group (LG), where SFHs of dwarf galaxies can be derived using the Color-Magnitude Diagram (CMD) method \citep{tol09}. The primary objective of this paper is to investigate the origin of dS0 galaxies by analyzing their SFHs. It is well-established that stellar mass plays a pivotal role in driving star formation in galaxies, resulting in the earlier formation of stars in more massive galaxies compared to less massive ones \citep{cow96, gav06, del07, san09}. The relationship between star formation rates (SFRs) and stellar mass is also well-documented \citep{bri04, dad07, noe07, sal07, rod11, sob14, spe14}.

The influence of the environment on the SFHs of galaxies has been studied extensively for bright galaxies \citep{spi51, oem74, dre80, kau04, wan22, per23}, as well as for early-type dwarfs, specifically dSphs and dEs \citep{seo23}. In this research, we focus on the SFHs of dS0s, paying special attention to the dependence of the cumulative star formation histories (cSFHs) on the physical and morphological properties of dS0s, as well as their environment. To achieve this, we derive the stellar populations by applying the population synthesis code STARLIGHT \citep{cidF05} to the spectra of these galaxies observed by the SDSS.

The structure is organized as follows. In Section 2, we outline the process of selecting sample galaxies and provide a brief overview of the method for analyzing spectra using STARLIGHT. Section 3 presents the SFHs of dS0s, focusing on luminosity and mass fractions. In Section 4, we present the cSFHs obtained from our analysis. Finally, in Section 5, we conclude the study by discussing our findings.

\section{Data and Method}
\subsection{Data}

We utilized the SDSS spectra of 148 dS0 galaxies, all of which are listed in the CVCG. While the CVCG initially includes 154 dS0 galaxies, we were able to obtain spectra for 148 dS0s from the SDSS Data Release 7 (DR7). It's worth noting that the CVCG encompasses a total of 5,638 galaxies situated in the local universe. Furthermore, the CVCG is nearly complete for galaxies brighter than $r=17.77$ in the regions covered by the SDSS. The signal-to-noise ratio (S/N) of their spectra mostly falls within the range of 10 to 30. The detailed classification of dS0 galaxies was provided by \citet{ann15}, who distinguished dS0s with spiral arm features as peculiar dS0s, and also differentiated between nucleated and non-nucleated dS0s. Among the 148 dS0 galaxies, 30 exhibit spiral arms in the outer regions, constituting roughly 20\% of the dS0 population. Concerning nucleation, the number of nucleated dS0 galaxies is comparable to that of non-nucleated dS0s.


\begin{figure}
	\centering
	\includegraphics[width=0.45\textwidth]{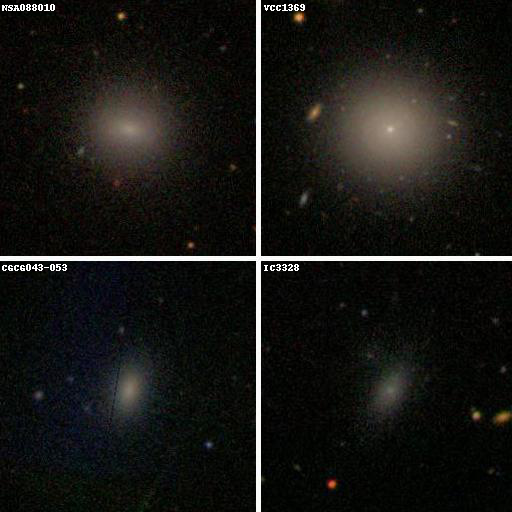}
	\caption{SDSS color images of four dS0 galaxies: NSA 088010, VCC 1369, IC 3328, and CGCG 043-053. The panels are arranged clockwise from the upper left. Galaxies in the upper row lack spiral arms, while those in the lower row exhibit them in the outer regions. Left panels show non-nucleated dS0s, and right panels depict nucleated dS0s. The box size in one dimension is $\sim$50 arcsec. North is at the top, and East is to the left.}
\end{figure}

Figure 1 presents color images of four dS0 galaxies: two without spiral arms (upper row) and two with spiral arms (lower row). The right panels show nucleated dS0s, while the left panels display non-nucleated dS0s. A notable observation from these images is the presence of a lens-like component in the inner regions of the dS0s, a distinctive feature.

As illustrated in Figure 2 where we can see the distribution of all CVCG galaxies in the $M_{r}$ versus $u-r$ color-magnitude diagram, dS0s with spiral arm features appear slightly fainter compared to those without such features. It's worth mentioning that nearly all dS0s with spiral arm features are fainter than $M_{r} =-16$. Furthermore, the majority of dS0s are situated along the red sequence \citep{str01}, if we consider the photometric error of $u-r$ color as $0.12 \pm0.08$. The error of $M_{r}$ is mostly due to distance error ($\delta D$) and we assumed $\delta D/D = 0.1$ to derive the error of $M_{r}$ in Figure 2. 

The SDSS spectra were captured using fibers with a diameter of 3 arcsec, positioned at the focal plane of the 2.5-meter telescope located at the Apache Point Observatory. The SDSS spectrograph incorporates 320 fibers, and the exposures lasted for 45 minutes or more, ensuring a fiducial signal-to-noise ratio. The spectra encompass a wavelength range spanning from 3800 to 9200 \AA, with an average spectral resolution of $\lambda/\Delta \lambda \sim 1800$. The wavelength and flux values have been calibrated using the pipeline developed by the SDSS team. We have also incorporated observational data from \citet{ann15}, which includes information such as distance, luminosity (M$_{r}$), color ($u-r$), morphological type, as well as coordinates and redshift.

\begin{figure}
	\centering
	\includegraphics[width=0.4\textwidth]{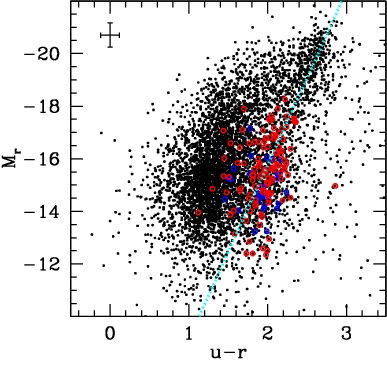}
	\caption{Color-magnitude diagram of CVCG galaxies. Blue circles represent dS0s with spiral arms, and red circles represent those without. The cyan dotted line indicates the red sequence from the CVCG catalog. Typical photometric error is shown in the upper left corner.
	}
\end{figure}

\subsection{Method}

We utilized the STARLIGHT code \citep{cidF05} to analyze the SDSS spectra of 148 dS0s. STARLIGHT allows us to determine the most likely combination of stellar populations based on their ages and metallicities. The STARLIGHT code has been thoroughly documented by \citet{cidF04} and \citet{cidF05}, and it has been applied in a number of studies including the recent studies such as \citet{rif21} and \citet{seo23}.

STARLIGHT fits the observed spectrum by generating a model spectrum derived from population synthesis models based on simple stellar populations (SSPs) from \citet{bc03}. These models incorporate six metallicities (Z = 0.0001, 0.0004, 0.004, 0.008, 0.02, and 0.05) and 25 age bins ranging from log (age)= 6 to 10.215 to represent the spectral evolution of SSPs. The stellar populations encompass stars with lower and upper mass cut-offs of $m_{L} = 0.1~\rm{M}_{\odot}$ and $m_{U} = 100~\rm{M}_{\odot}$. \citet{bc03} utilized basic stellar evolutionary tracks from the Padova groups, complemented by the Geneva groups' data, along with the STELIB library \citep{LeB03}.
	
In the STARLIGHT framework, the model spectrum $M_{\lambda}$ is represented as
\begin{equation}
  M_{\lambda}=M_{\lambda_{0}}(\sum_{j=1}^{N_{\ast}} x_{j} b_{j, \lambda} r_{\lambda}) \otimes G(v_{\ast},\sigma_{\ast})
\end{equation}

\noindent{where $b_{j, \lambda}$ is the $j-$th SSP spectrum normalized at $\lambda_{0}$,
$r_{\lambda}=10^{-0.4(A_{\lambda}-A_{\lambda_{0}})}$, $M_{\lambda_{0}}$ is
the synthetic flux at the normalization wavelength $\lambda_{0}$, and $x_{j}$ is
the fractional contribution of the SSP for $j-$th population that has
age $t_{j}$ and metallicity Z$_{j}$.The stellar motion projected on the line-of-sight is modeled by a Gaussian distribution (G) centered on the galaxy radial velocity $v$ with velocity dispersion $\sigma$. Extinction due to foreground dust is taken into account using the V-band extinction A$_{V}$ and the reddening law of \citet{car80}. We calculated A$_{V}$ using E(B-V) obtained from the dustmaps \citep{sch98}. The best fitting model is determined by
selecting a model that minimizes the $\Xi^{2}=\Sigma[(O_{\lambda}-M_{\lambda})w_{\lambda}]^{2}$ where $O_{\lambda}$ is observed spectrum and $w_{\lambda}$ is the inverse of error applied (see \citet{cidF04} for a detailed description).

\begin{figure}
	\centering
	\includegraphics[width=0.45\textwidth]{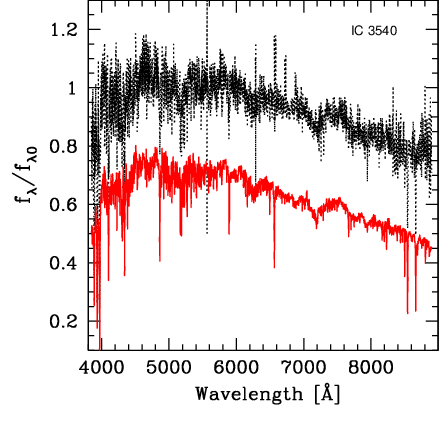}
	\caption{Normalized flux versus wavelength. Black line shows the observed spectrum, and the red line (offset by -0.3 for clarity) represents the best-fitting model spectrum from STARLIGHT. Both fluxes are normalized at $\lambda = 4020 \text{\AA} $.}
\end{figure}

We resampled the observed spectrum with a sampling width of $\delta\lambda$=1\AA,  after correcting the interstellar reddening and redshift following \citet{seo23}. In STARLIGHT, the luminosity fraction ($x_{j}$) and mass fraction ($\mu_{j}$) for the $j$-th stellar population are divided into six stellar
metallicities (Z=0.0001, 0.0004, 0.004, 0.008, 0.02, and 0.05), It assumes [$\alpha$/Fe]=0. The STARLIGHT output provides two mass fractions, $\mu_{ini}$ and $\mu_{cor}$, which represent initial mass and mass corrected for the mass returned to the interstellar medium, respectively. We used the $\mu_{cor}$ for the mass fraction of stellar populations while we used the $\mu_{ini}$ to derive the star formation rates.


\subsection{Uncertainties}

Figure 3 shows an example of population synthesis carried out using STARLIGHT. The model spectrum closely matches the details of the observed spectrum of IC 3540, a dwarf lenticular galaxy in the Virgo cluster. However, it is not a mathematical solution but a statistical solution, implying uncertainties in the derived physical parameters, such as stellar mass, mean age, and mean metallicity, are inevitable. Several factors introduce errors in the products of population synthesis models.

\begin{figure}
	\centering
	\includegraphics[width=0.45\textwidth]{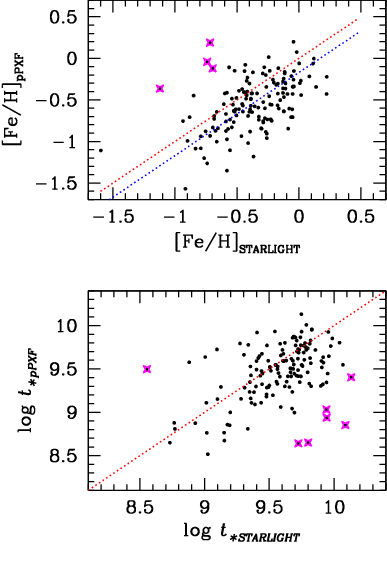}
	\caption{Comparison of mean ages and metallicities derived from STARLIGHT and pPXF. We plot the one-to-one relation as red dotted lines. Outliers, masked in the derivation of rms errors, are indicated by 'x' symbols (magenta). The blue dotted line is offset by -0.17 dex from the one-to-one line.}
\end{figure}

Noise in the observed spectrum is the primary source of errors. \citet{seo23} performed a mock test to examine the effect of random noise on the mean age and metallicity of 434 early-type dwarfs. They showed correlations between the mean metallicities and ages of stellar populations derived from the mock spectra and those of models, alongside the root mean square (rms) error of the mean metallicities. We used this data (their Figure 4) to calculate the rms errors of the mean ages derived from STARLIGHT. They range from 0.1 dex for spectra with signal-to-noise ratio ($S/N$) of 30 to 0.2 dex for spectra with $S/N=5$.
Our estimate of uncertainty in the mean age agrees with that of \citet{cidF05}. Similar uncertainty was also reported by \citet{mag15}, who noted that the uncertainty in log(age) for STARLIGHT is roughly the same as that of other nonparametric population synthesis methods. The lower limit of uncertainty in log (age) is set by the grid of isochrones in the base models. However, uncertainty of $\lesssim 0.2$ dex in log (age) is internal. There are biases resulting from the population synthesis models, which use different sets of SSPs as well as different numerical methods. The model dependence of the derived physical parameters was discussed by \citet{cidF14}, who showed $\sim$0.2 dex difference in mass, mean age, and metallicity among different base models.  

Here we present a comparision of the mean ages and metallicities derived from STARLIGHT and pPXF \citep{cap04}. The mean ages and metallicities are calculated using luminosity-weighted values through the following equations,

\begin{equation}
	\log t_{\ast} = \sum_{j=1}^{N_{\ast}} L_{j} \log t_{j}
\end{equation}

\begin{equation}
	Z = \sum_{j=1}^{N_{\ast}} L_{j} Z_{j}
\end{equation}

\noindent{where $L_{j}$ is the fractional luminosity contribution of $j$-th population and $\log t_{j}$ and $Z_{j}$ are its stellar age and metallicity, respectively.  } As shown in Figure 4,  the two datasets exhibit a correlation with significant scatter between them. The rms errors for both parameters are approximately $0.3$ dex. Moreover, the mean metallicities (upper panel) derived from pPXF deviate systematically from the one-to-one relation by -0.17 dex. However, the large rms errors are expected because, as described above, the rms error due to noise in the observed spectra amounts to approximately $0.1$ dex, and the rms error originating from the differences in the base models and calculation methods is about $0.2$ dex \citep{cidF14, mag15}. Thus, it appears better to treat higher-order physical parameter such as SFHs statistically, although physical parameters of individual galaxies still provide important information about their SFHs.

\section{Star Formation History}
\subsection{Luminosity and Mass Fraction}

Figure 5 shows the average luminosity and mass fractions of stellar populations in dS0 galaxies as a function of stellar age. We applied sigma clipping to the luminosity and mass fractions, excluding galaxies that deviated more than 3 standard deviations from the mean.  Fewer than 3\% of the galaxies were excluded. The resolution of stellar ages in Figure 5 reflects the age grid used in STARLIGHT. The age grid intervals are not uniform, ranging mostly from  0.1 dex and 0.2 dex with smaller interval for old isochrones. As previously mentioned, the age resolution is at best 0.1 dex and can be worse for spectra with $S/N$ less than 30. 

One prominent characteristic of the SFHs of dS0 galaxies is the presence of multiple bursts of star formation. The first period of active star formation began with an explosive burst at $\sim 14$ Gyr ago and concluded around 6 Gyr ago. There was a gradual decrease in star formation during this first period from 14 Gyr ago to 5 $\sim6$ Gyr ago. In contrast, the second period of active star formation exhibited a peak at $\sim$2.5 Gyr ago, spanning from around 6 Gyr ago to around 1 Gyr ago. Subsequently, star formation activity significantly decreased, reaching a nearly quiescent state. This pattern, characterized by initial bursts followed by a second peak and then near-quiescent star formation, is prevalent among dS0 galaxies, with $\sim$70\% exhibiting the initial burst and $\sim$85\% the second peak. 
Intriguingly, a minor resurgence of star formation activity occurred at $\sim$0.1 Gyr ago, albeit at significantly reduced intensity.  Also, the period of quenched star formation between $\sim$1 Gyr ago and $\sim$0.1 Gyr ago, represents a common feature observed in the SFHs of early-type dwarf galaxies within the LG \citep{tol09}.

Notably, around 60\% of the stellar mass formed during the first star formation period, and by the end of the second period, this value exceeded 99.8\%. Consequently, star formation in terms of stellar mass essentially terminated at roughly 1 Gyr ago. However, star formation continues to produce young, massive stars at a low level, as evidenced by the luminosity. The SFHs of dS0 galaxies exhibit striking similarities to the SFHs of dEs \citep{seo23}. This suggests that there are commonalities in the star formation histories of these two classes of early-type dwarf galaxies.

In terms of luminosity fractions, the initial burst of dS0 galaxies which contributes $\sim$10\% of the present luminosity is much weaker than the strongest burst at $\sim$2.5 Gyr ago and the stars formed in the first period of star formation contribute $\sim$30\% of the present luminosity in total. However, we consider the first period as the dominant phase of star formation because it produced around 60\% of the stellar mass. Furthermore, over half of the total stellar mass originates from stars older than roughly 10 Gyr.  The oldest stars, formed around 14 Gyr ago, currently comprise about 27\% of the total stellar mass and contribute around 10\% of the total luminosity. As described below, these oldest stars are primarily metal-poor (Z=0.0004).

Following the second star formation period, dS0 galaxies enter a phase of quenched star formation that lasts until roughly 0.1 Gyr ago. This is followed by a renewed increase in star formation activity that continues to the present day. However, the stars formed during this recent period contribute less than 0.3\% to the total stellar mass, with a total luminosity fraction of about 20\%. The near-complete absence of star formation between roughly 1 Gyr ago and 0.1 Gyr ago explains the lack of intermediate-age stars older than 0.1 Gyr. This suppression of star formation, characterized by a dearth of intermediate-aged stars, is also observed in some early-type dwarfs in the LG \citep{wei14a}, as well as in dSphs and dEs in the local universe \citep{seo23}. Stellar feedback mechanisms, such as supernova explosions and stellar winds, are believed to be responsible for quenching star formation. These processes heat the cold gas within the galaxies, making it easier to be expelled into the halos of early-type dwarfs. 

\begin{figure}
	\centering
    \includegraphics[width=0.45\textwidth]{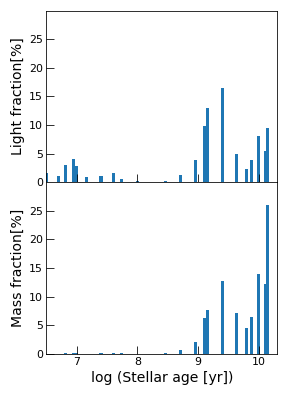}	
	\caption{Mean luminosity and mass fractions as a function of stellar age for 148 dS0 galaxies. The upper panel shows the luminosity fraction, while the lower panel shows the mass fraction.}
\end{figure}

\subsection{Metallicity Dependent Mass Fraction}

Figure 6 presents the average mass fractions of stellar populations within 148 dS0 galaxies, categorized into six metallicity groups, as a function of stellar age. Similar to Figure 5, we excluded outliers using sigma clipping before calculating average mass fractions. The metallicities considered are Z=0.0001, Z=0.0004, Z=0.004, Z=0.008, Z=0.02, and Z=0.05. The most pronounced feature of the dS0 SFHs is the dearth of extremely metal-poor stars (Z=0.0001) formed before $\sim$10 Gyr ago. Among the stars formed before this time, contributing $\sim$50\% of the current stellar mass, over $\sim$80\% are metal-poor. The remaining stars have metallicities of Z=0.004, 0.008, 0.02. and 0.05, each contributibg at most $\sim$5\%. Interestingly, stars formed during initial burst at $\sim$14 Gyr ago are also predominantly (80\%) metal-poor stars. This lack of extremely metal-poor stars formed during the first star formation period is also observed in dSphs and dEs \citep{seo23}. This is thought to be a consequence of pre-enrichment during the re-ionization era.

\begin{figure}
\centering
\includegraphics[width=0.45\textwidth]{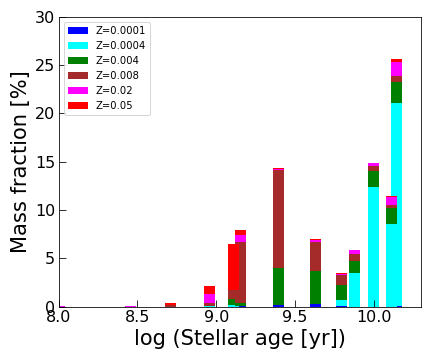}
\caption{Mean mass fractions of stellar populations as a function of stellar age for 148 dS0 galaxies. The stellar populations are divided by their metallicities.}
\end{figure}

The chemical evolution in dS0 galaxies during the second period of star formation closely parallels that of dEs \citep{seo23}. Specifically, it starts with the formation of intermediate metallicities (Z=0.004 and Z=0.008) and ends with the emergence of extremely metal-rich stars. Within the second period of star formation, the predominant metallicity is the intermediate metallicity of Z=0.008, accounting for more than half of the stars formed during this phase. This high prevalence of stars with Z=0.008 is attributed to the rapid enrichment process in dS0 galaxies.

Due to this swift enrichment, extremely metal-rich stars are produced in the latter stages of the second period of star formation. In fact, all generations of stars formed during the first and second periods of star formation exhibit some extremely metal-rich stars, although their contribution to the current stellar mass is relatively small, with the exception of stars formed at $\sim$1.2 Gyr ago. 

The occurrence of reduced star formation in dS0 galaxies after initial burst of star formation could be attributed to the explosive starbursts that took place during the initial phase. The tremendous energy released during the starbursts efficiently expelled the remaining gas into the galactic halo. However, this gas is not lost forever. Over time, some of the gas retained within the halo, eventually falls back, leading to star formation after $\sim$6 Gyr ago, i.e., initiating a second period of star formation. Similar behavior has also been observed in dEs as well \citep{seo23}.

\begin{figure}
	\centering
	\includegraphics[width=0.4\textwidth]{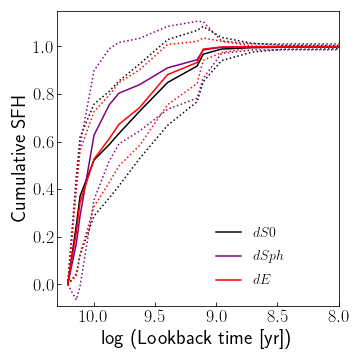}
	\caption{Mean cSFHs of dS0s compared with those of dEs and dSphs \citep{seo23}. Solid lines represent the mean cSFHs and dotted lines designate $\pm1\sigma$ boundaries. 
	}
	
\end{figure}
%

\section{Cumulative Star Formation History}

\subsection{Mean cSFHs}

As shown in Figures 5 and 6, the mean SFH of dS0 galaxies varies significantly since the beginning of star formation at the lookback time of $\sim$14 Gyr. To understand the mass assembly history of dS0 galaxies more clearly we examine the cumulative star formation histories (cSFHs) of dS0s in detail below. We examine the mean cSFH of dS0 galaxies rather than cSFH of individual galaxies because, as shown in the appendix, the cSFH of individual galaxy is very diverse. We derived a mean cSFH of dS0 galaxies by applying 3$\sigma$ clipping. The mean cSFH could help to identify overarching trends and patterns that apply broadly to this class of galaxies, shedding light on their evolutionary processes. For comparison with other early-type galaxies, dSphs and dEs which were studied by \citet{seo23}, we plotted the cSFHs of these types together in Figure 7.

The general behaviour of the cSFHs of dS0s is characterized by a rapid increase of cSFH due to bursts of star formation occurring before the lookback time of 10 Gyr, contributing to $\sim$50\% of the present stellar mass. Subsequently, star formation activity significantly decreases resulting in slow increase until the lookback time of $\sim$0.3 Gyr. There is no discernible increase in cSFH due to the almost complete quenching of star formation from the lookback time of $\sim$0.3 Gyr. 

The cSFHs of dEs closely resemble those of dS0s, while dSphs exhibit significantly different cSFH patterns. \citet{seo23} attributed the difference in the cSFHs of dEs and dSphs to their origins. The majority of dSphs are primordial galaxies while dEs, at least significant fraction of them, are transformed from late-type galaxies. The similarity between dS0s and dEs suggests that dS0s may also originate from late-type galaxies. The morphological properties which are characterised by the presence of central lens component strongly supports the transformation origin of dS0 galaxies. It is worth to note that, VCC \citep{bin85} and EVCC \citep{kim14} do not differentiate between dSphs and dEs. However, as shown in Figure 7, the cSFHs of these two types are signicantly different. The similarity in cSFHs between dS0s and dEs suggests potential commonalities in their evolutionary history. This contrast and comparison between the cSFHs of three sub-types of early-type dwarfs may provide insights into the formation and evolution of early-type dwarf galaxies within the broader context of galaxy evolution.

\subsection{Dependence on Physical and Environmental Properties}

\subsubsection{Morphology}

There are two distinct morphological features associated with dS0  galaxies, outer spiral arms and nucleation. Outer spiral arms are unique features observed specifically in dS0 galaxies, while nucleation is a common trait among early-type dwarfs. Despite nucleation being a prevalent characteristic in early-type dwarfs (encompassing dS0s, dEs, and dSphs), the fraction of nucleated dS0 galaxies is smaller than those of dEs and dSphs, both of which exhibit nucleation in $\sim$85\% of their populations. This discrepancy is likely due to the presence of the central lens component, which distinguishes dS0 galaxies as a distinct sub-type within the category of early-type dwarfs. The fraction of dS0 galaxies that possess outer spiral arms is 0.2.

Figure 8 illustrates the mean cSFHs of dS0 galaxies as a function of lookback time, categorized based on the presence or absence of outer spiral arms and nucleation. In the CVCG, dS0 galaxies with outer spiral arms are classified as dS0${p}$, while those with nucleation are labeled as dS0${n}$. In the left panel of Figure 8, we present the mean cSFHs of these 30 dS0${p}$ galaxies along with the 118 remaining dS0 galaxies obtained via 3$\sigma$ clipping. There is no significant difference in the cSFHs between dS0${p}$ galaxies and dS0 galaxies. A marginal difference exists between the two samples only between the lookback time of $\sim$10 Gyr and $\sim$2.5 Gyr. Given the lack of significant differences between the cSFHs of dS0$_{p}$ galaxies and dS0 galaxies, we will combine the two samples and refer to them collectively as dS0s in the subsequent analysis. In the right panel of Figure 8, we present the mean cSFHs of dS0s, divided into those with and without nucleation. Similar to the cSFHs of dS0 galaxies categorized by the presence or absence of spiral arm features, the cSFHs of dS0 galaxies with and without nucleation are quite similar. However, a slightly more rapid star formation phase is observed between the lookback time of $\sim$10 Gyr and $\sim$1 Gyr in the cSFH of dS0 galaxies that lack nucleation. Consequently, the presence or absence of additional morphological features, such as outer spiral arms and nucleation does not significantly influence the cSFHs of dS0 galaxies.

\begin{figure}
	\centering
	\includegraphics[width=0.45\textwidth]{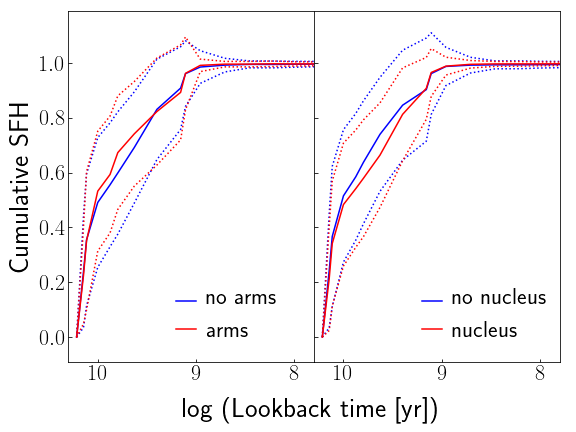}
	\caption{Mean cSFHs of early-type dwarfs with and without spiral arms (left panel) and nucleation (right panel). The line styles are the same as those in Figure 7. }
	
\end{figure}

\subsubsection{Stellar Mass}

We determined the stellar mass of a galaxy from the model flux fitted by STARLIGHT by using the galaxy distance in the CVCG. Since the flux obtained through the $3^{\prime\prime}$ fiber which covers a small fraction of a galaxy image, we applied an aperture correction ($AC$) calculated as
\begin{equation}
	AC= \frac{2\times \int_0^{R_{e}} 2\pi f(r) r dr} { \int_0^{R_{f}} 2\pi f(r) r dr}
\end{equation}
\noindent{where	$R_{e}$ and $R_{f}$ are the effective radius and the fiber radius, respectively, and {\it{f(r)}} is the S\'{e}rsic profile \citep{ser68}. We assumed an  axis-symmetric luminosity distribution and used the fact that the luminosity within $R_{e}$ is half of the total luminosity. We use the $R_{e}$ and  S\'{e}rsic index ($n$) that were determined by \citet{seo22}. For galaxies with unknown $R_{e}$ and $n$, we use the mean $R_{e}$ and $n$ derived for dS0s. The average $R_{f}/R_{e} = 0.23 \pm 0.18$ and it ranges from 0.05 to 0.6.}  The number of galaxies with unknown S\'{e}rsic parameters is $\sim10\%$ of the sample galaxies. The stellar masses of dS0 galaxies were found to fall within the range of $\sim$5$\times10^{6}$ M$_{\odot}$  to  $\sim$$10^{10}$ M$_{\odot}$. 

\begin{figure}
	\centering
	\includegraphics[width=0.4\textwidth]{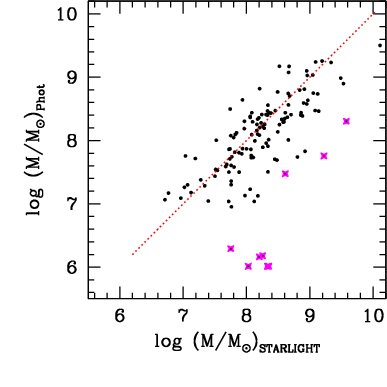}
	
	\caption{Comparision between stellar masses derived in this study and those from photometry based on SDSS DR12. One-to-one relation is give by a red dotted line. The outliers, not used in the calculation of rms error, are marked by 'x' symbols (magenta).}
	
\end{figure}

Figure 9 shows a comparison between the present estimates of the total stellar mass and those from SDSS DR12. The total stellar masses provided by SDSS DR12 were derived from multi-band photometric images following \citet{kau03} and \citet{bri04}. Approximately $10\%$ of the sample galaxies are missing in the stellar mass table of SDSS DR12. There is a pretty good correlation between the two sets of stellar masses because they appear to scatter around the one-to-one relation. The rms in the correlation is $0.4$ dex, which is larger than that of the correlation in the age estimates, $0.3$ dex. The rms error of 0.4 dex is in a good agreement with the rms reported by \citet{cidF05}. In the derivation of rms error we exclude outliers, indicated by 'x' symbols (cyan) in Figures 4 and 9.

Given the current consensus in the community that the mass of a galaxy plays a crucial role in star formation, particularly in early-type dwarf galaxies, we divided the dS0s into two groups based on their stellar mass. The high mass group comprises the dS0s with stellar masses greater than the median mass plus one standard deviation ($\sigma$), while the low mass group includes the dS0s with stellar masses smaller than the median mass minus one standard deviation ($\sigma$).

In Figure 10, we present the mean cSFHs of dS0s along with the $1\sigma$ boundaries, segregated into these two groups, as a function of lookback time. We calculated the mean cSFHs by applying 3$\sigma$ clipping. As depicted in Figure 10, the cSFHs of dS0s exhibit significant differences when grouped by stellar mass. The average cSFH of the high mass group shows a more rapid increase compared to that of the low mass group. Notably, the cSFH of the low mass group lies outside the $1\sigma$ boundary of the high mass group for lookback times between $\sim$3 Gyr and $\sim$1.6 Gyr. This suggests a strong stellar mass dependence in the cSFHs of dS0s, which appears to be slightly more pronounced than the dependencies observed in dSphs and dEs, as reported in \citet{seo23}.

	\begin{figure}
		\centering
		\includegraphics[width=0.4\textwidth]{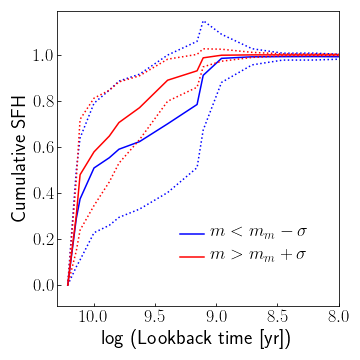} 
		
		\caption{Mean cSFHs of dS0s grouped by stellar mass. Galaxies in the lower mass group have stellar mass smaller than $m_{m}$ (median mass) - $\sigma$ while those in the high mass group have stellar mass larger than $m_{m}$ + $\sigma$. The line styles are the same as those in Figure 7. }
		
	\end{figure}

\subsubsection{Background density}

The SFHs of galaxies are known to be influenced by their environment \citep[e.g.,][]{kau04}. In this study, we employed the background density ($\Sigma$) as a measure of galaxy environment. There are several methods to calculate background density \citep[see][for details]{mul12}, but for this analysis, we utilized the $n$-th nearest neighbor method with $n=5$. This method necessitates two constraints for selecting neighboring galaxies: the linking velocity ($\Delta V^{\ast}$) and the limiting magnitude ($M_{lim}$). 
We set the linking velocity $\Delta V^{\ast}$ to be 500 km s$^{-1}$ and the limiting magnitude as $M_{r} = -15.2$ for local effect and  $M_{r} = -20.6$ for global effect. They are the r-band absolute magnitudes. The former corresponds to the luminosity to define a volume-limited sample for galaxies with redshifts less than $z=0.01$ and the latter is the $M^{\ast}$ defined in the luminosity function of the local universe \citep{ann15}. The choice of $\Delta V^{\ast}$ as 500 km s$^{-1}$ is motivated by the peculiar velocity of the local universe \citep{pee79} as well as a detailed analysis of 
$\Delta V^{\ast}$ by \citet{ann14}.

We calculated the background density $\Sigma$ using the following equation,
\begin{equation}
	\Sigma=\frac{n}{{\pi{r_{p}}^{2}}}
\end{equation}
where $r_{p}$ is the projected distance to the $n$th nearest neighbor
galaxy. We normalized $\Sigma$ using the mean background
density ($\bar{\Sigma}$) of the galaxies in the local universe within $z = 0.01$. In the calculation of $\Sigma$, we utilized the galaxy distances from the CVCG.

\begin{figure}
	\centering
	\includegraphics[width=0.4\textwidth]{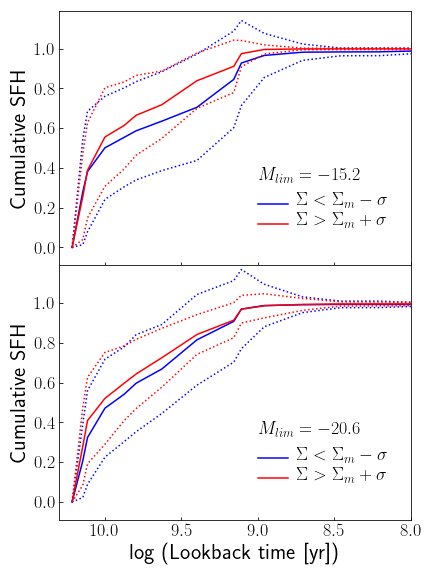}
	\caption{Mean cSFHs of dS0s grouped by the background density. We present the background density constrained by $M_{lim}=-15.2$ in the upper panel and that with $M_{lim}=-20.6$ in the lower panel. Here $\Sigma_{m}$ and $\sigma$ are the median background density and the starndard deviation, respectively. The line styles are the same as those in Figure 7.
	}
	
\end{figure}

Figure 11 presents the cSFHs of dS0s grouped by their background density ($\Sigma$). Similar to the approach used for physical parameters, we divided the galaxies into two groups based on the median and standard deviation ($\sigma$) of the background density. The low density group consists of galaxies with $\Sigma < \Sigma_{m} - \sigma$, while the high density group includes galaxies with $\Sigma > \Sigma_{m} + \sigma$, where $\Sigma_{m}$ represents the median of the background density.

It is apparent that a distinction exists between the two groups in their cSFHs when grouped by background density, as constrained by $M_{r} = -15.2$. However, there is not a substantial difference observed for the two groups constrained by $M_r = -20.6$. This suggests that star formation in dS0s is likely more influenced by the local environment than by the global environment associated with large-scale structures. The most pronounced difference in the cSFHs of the two density groups occurs around the lookback time of 2 Gyr.

During the second period of star formation, where intermediate-age stellar populations play a significant role, the background density appears to have a more pronounced impact. High-density regions are more likely to witness star formation during this phase. It's important to note that the stars formed during the second period of star formation do not originate from primordial gas but rather from enriched gas that has been ejected and subsequently falls back into the galaxy after a period of highly reduced star formation. In this scenario, a denser environment may be advantageous in retaining the ejected gas for a longer duration, thereby facilitating its involvement in subsequent star formation episodes.

\begin{figure}
	\centering
	\includegraphics[width=0.42\textwidth]{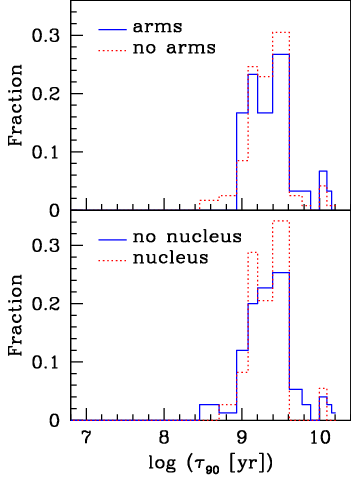}
	\caption{Frequency distributions of quenching time $\tau_{90}$. The upper panel shows the quenching times grouped by outer spiral arms, while the lower panel shows the quenching times grouped by nucleation. }
\end{figure}

\subsection{Quenching Epoch}

Figure 12 displays the distribution of quenching time ($\tau_{90}$) for 148 dS0s categorized by the presence of spiral arms and nucleation. The quenching time ($\tau_{90}$), introduced by \citet{wei14a, wei14b}, quantifies the cSFH of a galaxy. It represents the lookback time at which 90\% of the stellar mass has formed. The distribution of $\tau_{90}$ values is roughly Gaussian with tails at both ends. Notably, around 10\% of dS0s exhibit $\tau_{90}$ exceeding 10 Gyr. 

As anticipated due to slight variations in the SFHs of dS0s with different morphologies, there are discernible differences in $\tau_{90}$ values among dS0s with distinct morphologies. Concerning the presence of spiral arm features, no dS0 galaxy with spiral arms has a $\log(\tau_{90}$) less than 9, while a small fraction of dS0s without spiral arms quenched at lookback time of $\lesssim1$ Gyr. However, no significant difference in late-time quenching is observed between dS0s with and without nucleation.

In Figure 13, we present the distribution of $\tau_{90}$ for all dS0s, along with those of dSphs and dEs for comparison. As discussed in \citet{seo23}, the extremely early quenching observed in some early-type dwarf galaxies is likely due to their low mass. Galaxies with lower mass tend to quench at earlier stages \citep{dig19, gar19, jos21}. This aligns with the fact that dSphs have a larger fraction of early-quenched galaxies compared to dEs and dS0s. The $\tau_{90}$ distribution of dS0s bears a greater resemblance to that of dEs than to dSphs. Early-type dwarf galaxies with $\tau_{90}$ exceeding 10 Gyr represent genuine primordial objects. 

\section{Discussion and Conclusions}

This analysis of the star formation histories (SFHs) of 148 dS0s, using SDSS spectra and STARLIGHT \citep{cidF05}, revealed distinct characteristics. A prominent initial burst of star formation occurred at a lookback time of around 14 Gyr, followed by subsequent bursts that peaked at around 10 Gyr and 2.5 Gyr.  The 10 Gyr peak was dominated by metal-poor stars, while the 2.5 Gyr peak was rich in stars with intermediate metallicity. Interestingly, extremely metal-poor stars (Z=0.0001) are scarce during early galaxy formation, suggesting potential pre-enrichment by Population III stars during the re-ionization era.

Our investigation further suggests that stellar feedback plays a crucial role in the early cessation of star formation in many dS0s, resulting in gaps in their SFHs. This early quenching effect is also observed in dSphs and dEs \citep{seo23}. However, among the three types of early-type dwarf galaxies, early quenching is most pronounced in dSphs, while dS0s exhibit an intermediate level compared to dSphs and dEs.

\begin{figure}
	\centering
	\includegraphics[width=0.42\textwidth]{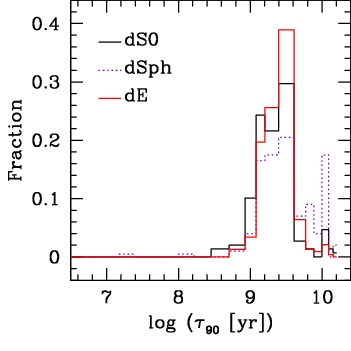}
	\caption{Distributions of quenching time $\tau_{90}$ for our dS0 sample, along with those of dSphs and dEs for comparison. The data for dSphs and dEs were obtained from \citet{seo23}. 	}
\end{figure}

The star formation activity within dS0s can be divided into three distinct periods. The first period exhibits two prominent peaks, one at the lookback time of $\sim$14 Gyr and the other at the lookback time of $\sim$10 Gyr. The first peak marks the onset of  a major starburst phase, responsible for 27\% of the present-day stellar mass in dS0 galaxies. The second peak represents another significant epoch for the formation of metal-poor stars.

The second period spans from between the lookback time of $\sim$6 Gyr to $\sim$1 Gyr, with the bulk of star formation occurring at the lookback time of $\sim$2.5 Gyr. This period predominantly gives rise to intermediate-metallicity stars. In dS0s, the stellar mass produced at the peak of the second period is comparable to that generated during the earlier peak at 10 Gyr. In the second period of star formation, the level of star formation activity is comparatively lower than that observed in the first period. However, stars with intermediate metallicity formed around the peak of the second period contribute $\sim$30\% of the current stellar mass. The third period of star formation recommences around the lookback times of $0.1$ Gyr after a phase of quiescent star formation. However, its contribution to the current stellar mass is  negligible.

We have undertaken an investigation into the cSFHs of dS0s, taking into account various factors including morphology, stellar mass, and background density. The cSFHs of dS0s reveal discernible dependencies on physical properties: stellar mass and local background density. In constrast, the presence or absence of morphological features such as outer spiral arms and nucleation do not exhibit statistical significance. Additionally, it is also apparent that dS0s appear to form stars more rapidly in high-density environments. 

The lack of significant dependence of cSFHs on morphological features, the outer spiral arms and nucleation, particularly during the early stages of galaxy formation, suggests that these morphological traits likely  developed after the initial period of star formation. This is supported by the observed differences in cSFHs that occur after the lookback time of $10$ Gyr. Additionally, the outer spiral arms have minimal influence on star formation in the central regions of these galaxies, as these features are typically located outside the central star-forming regions.

While the dependence of cSFHs on the background density ($\Sigma$) is weaker than stellar mass in dS0s, it remains a significant factor. This is because some dS0s might have transformed from late-type galaxies, often driven by environmental interactions. One indicator of this transformation is the presence of embedded structures in dS0 galaxies, reminiscent of late-type galaxies. A similar situation is observed in dEs, where a notable fraction of them also possess embedded disc features \citep{seo22}. We propose that the second peak in the SFHs of dS0s and dEs might be linked to this potential transformation. However, a significant fraction of dS0s exhibit early quenching ($\tau_{90}$ > 10 Gyr), suggesting a primordial origin. In contrast, dSphs, mostly considered primordial objects, lack this second peak. 

The scarcity of extremely metal-poor (Z = 0.0001) stars in dS0s hints at pre-enrichment of the gas that formed these galaxies. This  aligns with findings in dEs and dSphs \citep{seo23}, where the absence of such stars is attributed to the rapid metal enrichment by Population III supernovae. This lack of extremely metal-poor stars is consistent with observations across various environments, including the Milky Way halo stars, Local Group dwarf spheroidals \citep{hel06}, damped Ly${\alpha}$ absorption systems \citep{wol98}, and the intergalactic medium towards quasars \citep{cow98}.Simulations \citep{wis12} also support pre-enrichment, potentially leading to metallicities as high as Z = 0.0004 for the oldest stars.

\section*{Acknowledgements}

HBA thanks the anonymous reviewer for their valuable comments, which significantly improved the paper. He also thanks Dr. Roberto Cid Fernandes for providing STARLIGHT.



\section*{DATA AVAILABILITY} 

We provide the basic output of STARLIGHT as supplemented materials
and additional data are available upon request.


\appendix

\section{Cumulative Star Formation Histories of 148 dS0 galaxies}

Figure A1 shows the cSFHs of the 148 dS0s. A small fraction ($\lesssim 3\%$) of dS0s exhibit very early quenching, while a similar fraction of dS0s show very delayed star formation. The SFHs of the 148 dS0s are provided in tabular form as supplementary data.

\begin{figure}
	\centering
	\includegraphics[width=0.45\textwidth]{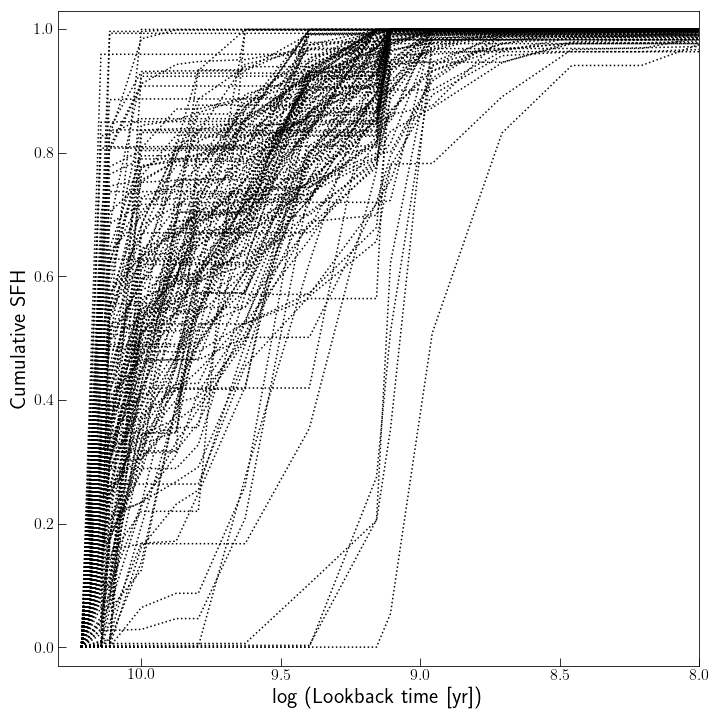}
	\caption{Cumulative SFHs of 148 dwarf lenticular galaxies.
	}
\end{figure}}

\bsp    
\label{lastpage}

\end{document}